\documentstyle[11pt,osid,amsmath,amssymb,graphicx]{article}

\title{Microscopic derivation of open quantum walk on two node graph}
\author{Ilya Sinayskiy 
		  \\{\footnotesize\it Quantum Research Group, School of Chemistry and Physics and National Institute for Theoretical Physics, University of KwaZulu-Natal, Westville Campus, Durban, South Africa\\ e-mail: sinayskiy@ukzn.ac.za}\\[2ex]
        Francesco Petruccione
                   \\{\footnotesize\it Quantum Research Group, School of Chemistry and Physics and National Institute for Theoretical Physics, University of KwaZulu-Natal, Westville Campus, Durban, South Africa\\ e-mail: petruccione@ukzn.ac.za} }

\begin{document}

\maketitle
\begin{abstract}
A microscopic derivation of an open quantum walk on a two node graph is presented. It is shown that for the considered microscopic model of the system-bath interaction the resulting quantum master equation takes the form of a generalised master equation. The explicit form of the ``quantum coin" operators is derived. The formalism is demonstrated for the example of a two-level system walking on a two-node graph. 
\end{abstract}

\section{Introduction}

Recently  Attal et al. introduced a new type of quantum walks, so called open quantum walks \cite{pla, JSP}. In an open quantum walk (OQW) the quantum walker is exclusively driven by the interaction with an environment. In this approach the dissipation is the factor which determines the dynamics of the quantum walker, as opposed to unitary quantum walk approaches \cite{kempe,qwqc1}, where the dissipative effects need to be minimised or eliminated \cite{dqw1,dqw2}. The formalism of OQWs rests upon the implementation of  completely positive trace preserving maps (CPTP). OQWs show rich dynamical behaviour and can be used to perform efficient dissipative quantum computation and state engineering \cite{pla, QIP}. The properties of the asymptotic probability distribution of the OQWs were analysed as well \cite{clt1,clt2, PhysSc}. However, so far no microscopic model of system-environment interaction which leads to OQWs has bee reported. In this paper we present a model of an open quantum system whose discrete time reduced dynamics is given by an OQW. 

The paper is organised as follows. In Section 2 we review the OQW formalism. In Section 3 we present a microscopic model of system environment interaction and derive the corresponding quantum Markov equation. In Section 4 we construct the explicit form of the OQW based on the generalised quantum master equation derived in Section 3. In Section 5 we present a simple example of the derivation and in Section 6  we conclude. 

\section{The formalism}

OQWs are defined as completely positive trace preserving maps. These CPTP-maps correspond to some dissipative processes which are driving the transition between the nodes of the graph. The underlying graph is defined by the set of vertices $\cal{V}$ with oriented edges $\{(i,j)\,;\ i,j\in\cal{V}\}$. The number of nodes is considered to be finite or countable infinite. The space of states corresponding to the dynamics on the graph denoted by $\begin{cal}K\end{cal}=\begin{mathbb}C\end{mathbb}^{\begin{cal}V\end{cal}}$. If $\cal{V}$ is an infinite countable set, the space of states $\cal{K}$ will be any separable Hilbert space with an orthonormal basis ${(| i\rangle)}_{i\in\cal{V}}$ indexed by $\cal{V}$. The internal degrees of freedom of the quantum walker, e.g., the polarisation, spin or $n$-energy levels, will be described by a separable Hilbert space $\cal{H}_N$ attached to each node. Any state of the walker will be described on the direct product of the Hilbert spaces $\cal{H}_N\otimes \cal{K}$.

To describe the dynamics of the internal degree of freedom of the walker for each edge $(i,j)$ we introduce  a bounded transition operator $B^i_j$ acting on $\cal{H}_N$. This transition operator describes the change in the internal degree of freedom of the quantum walker due to the "jump" from node $j$ to node $i$ (see Fig. 1). By choosing for each node $j$ that, 
\begin{equation}\label{eq1}
\sum_i {B^i_j}^\dag B^i_j= I,
\end{equation}
we guarantee, that for each node of the graph $j\in\mathcal{V}$ there is a corresponding CPTP map on the operators of $\mathcal{H}_N$:
\begin{equation}
\begin{mathcal}M\end{mathcal}_j(\tau)=\sum_i B^i_j \tau {B^i_j}^\dag.
\end{equation}

\begin{figure}
\begin{center}
\includegraphics[width= .6\linewidth]{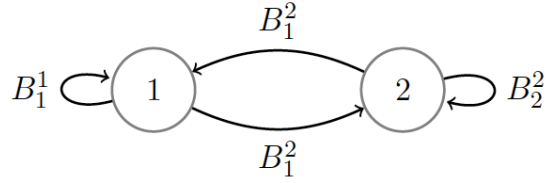}
\end{center}
\caption{A schematic representation of an open quantum walk on a 2-node graph. The operators $B_i^j$ $(i, j = 1,2)$ represent the ``quantum coin" operators of the walk.}
\end{figure}

The transition operators $B^i_j$ act only on the internal or ``quantum coin" Hilbert space $\mathcal{H}_N$ and do not perform transitions from node $j$ to node $i$. They can be easy dilated to operators $M^i_j$ acting on the total Hilbert space $\mathcal{H}\otimes\mathcal{K}$  in the following way
\begin{equation}
M^i_j=B^i_j\otimes | i\rangle\langle j|\,.
\end{equation}
It is clear that, if the condition expressed in Eq. (\ref{eq1}) is satisfied, then $\sum_{i,j} {M^i_j}^\dag M^i_j=1$ \cite{JSP}. This condition defines a CPTP map for density matrices on $\mathcal{H}_N\otimes\mathcal{K}$, i.e.,
\begin{equation}\label{OQW}
\begin{mathcal}M\end{mathcal}(\rho)=\sum_i\sum_j M^i_j\,\rho\, {M^i_j}^\dag.
\end{equation}
The CPTP-map $\begin{mathcal}M\end{mathcal}$ defines the discrete time \textit{Open Quantum Walk} \cite{pla, JSP}. One can see that for an arbitrary initial state the density matrix $\sum_{i,j} \rho_{i,j}\otimes| i\rangle\langle j|$ will take a diagonal form in position Hilbert space $\cal{K}$ after just one step of the OQW Eq. (\ref{OQW}),
\begin{eqnarray}
\nonumber
\begin{mathcal}M\end{mathcal}\left(\sum_{k,m} \rho_{k,m}\otimes|k\rangle\langle m|\right)&=&\sum_{i,j,k,m} B^i_j\otimes | i\rangle\langle j|\,\left(\rho_{k,m}\otimes|k\rangle\langle m|\right)\, {B^i_j}^\dag\otimes | j\rangle\langle i|\\\nonumber
&=&\sum_{i,j,k,m} B^i_j\rho_{k,m}{B^i_j}^\dag\otimes | i\rangle\langle i|\delta_{j,k}\delta_{j,m}\\
&=&\sum_i \left(\sum_j B^i_j\rho_{j,j}{B^i_j}^\dag\right)\otimes | i\rangle\langle i|.
\end{eqnarray}
We will assume that the initial state of the system has the form
\begin{equation}\label{rho}
\rho=\sum_i \rho_i\otimes | i\rangle\langle i|,
\end{equation}
with
\begin{equation}
\sum_i { \mathrm Tr}(\rho_i)=1.
\end{equation}
It is straightforward to give an explicit iteration formula for the OQW from step $n$ to step $n+1$ 
\begin{equation}\label{rho}
\rho^{[n+1]}=\begin{mathcal}M\end{mathcal}(\rho^{[n]})=\sum_i \rho_i^{[n+1]}\otimes | i\rangle\langle i|,
\end{equation}
where
\begin{equation}
\rho_i^{[n+1]}=\sum_j B^i_j \rho_j^{[n]}{B^i_j}^\dag.
\end{equation}

The above iteration formula gives a clear physical meaning to the CPTP-mapping that we introduced: the state of the system on  site $i$ is determined by the conditional shift from all connected nodes $j$, which are defined by the explicit form of the quantum coin operators $B_j^i$ and an internal state of the walker on the node $j$.  Also, one can see that $\mathrm{Tr}[\rho^{[n+1]}]=\sum_i\mathrm{Tr}[\rho_{i}^{[n+1]}]=1$.

As a particular example of an OQW we consider the simplest case of a walk on a 2-node graph (see Fig. 1). In this case the transition operators $B_i^j$ $(i,j=1,2)$ satisfy:
\begin{equation}
\label{eq:norm2}
B_1^{1\dag}B_1^1+B_1^{2\dag}B_1^2=I,\quad B_2^{2\dag}B_2^2+B_2^{1\dag}B_2^1=I.
\end{equation}
The state of the walker $\rho^{[n]}$ after $n$ steps is given by,
\begin{equation}
\rho^{[n]}=\rho_1^{[n]}\otimes|1\rangle\langle 1|+\rho_2^{[n]}\otimes|2\rangle\langle 2|,
\label{2node}
\end{equation}
where the form of the $\rho_i^{[n]}$ $(i=1,2)$ is found by iteration,
\begin{eqnarray}
\rho_1^{[n]}=B_1^1\rho_1^{[n-1]}B_1^{1\dag}+B_2^1\rho_2^{[n-1]}B_2^{1\dag},\\\nonumber
\rho_2^{[n]}=B_1^2\rho_1^{[n-1]}B_1^{2\dag}+B_2^2\rho_2^{[n-1]}B_2^{2\dag}.
\end{eqnarray}

\section{Microscopic model and Markov master equation}

As for arbitrary open quantum system, the total Hamiltonian of system and bath can be presented in the following form \cite{toqs},
\begin{equation}
H=H_S+H_B+H_{SB},
\end{equation}
%\begin{mathcal}\end{mathcal}
where $H_S$ is Hamiltonian of the system. This Hamiltonian describes the non-interacting free dynamics of two nodes on a two node graph ($H_S\in\begin{mathcal}B(\mathcal{H}_{N}\end{mathcal}\otimes\begin{mathbb}C\end{mathbb}^{2})$), 
\begin{equation}
H_S=\Omega_1\otimes|1\rangle\langle 1|+\Omega_2\otimes|2\rangle\langle 2|,
\end{equation}
where the Hamiltonians $\Omega_i$ ($i=1,2$) describe the dynamics of the internal degree of freedom of the walker on the node ($\Omega_i\in\begin{mathcal}B(\mathcal{H}_{N}\end{mathcal})$).
The bath is modelled as a set of harmonic oscillators, so the Hamiltonian $H_B$ of the bath is given by,
\begin{equation}
\label{eq:bath}
H_B=\sum_n \omega_n a^\dag_na_n,
\end{equation}
where $a^\dag_n$ and $a_n$ are creation and annihilation operators satisfying standard bosonic commutation relations and $\omega_n$ is the frequency of the corresponding harmonic oscillator.

The Hamiltonian of the system-bath interaction has the following form,
\begin{equation}
H_{SB}=A\otimes X\otimes B,
\end{equation}
where $A$ is a hermitian operator on the node Hilbert space ($\cal{H}_N$) which describes the change in the internal degree of freedom of the walker corresponding to the interaction with the environment. The operator $X$ describes the transition between nodes, $X=|1\rangle\langle 2|+|2\rangle\langle 1|$ and the operator $B=\sum_ng_na_n+g_n^*a_n^\dag$ describes the linear coupling of the walker to the  bath. 

Here we will assume that the Hamiltonians of the walker at the nodes are non-degenerate and can presented as,
\begin{equation}
\Omega_1=\sum_{i=1}^N \lambda_i |\lambda_i\rangle\langle \lambda_i|, \quad \Omega_2=\sum_{i=1}^N \xi_i |\xi_i\rangle\langle \xi_i|,
\end{equation}
where $\lambda_i$ and $\xi_i$ denote energy levels and $|\lambda_i\rangle$ and $|\xi_i\rangle$ the corresponding eigenvectors.

In the interaction picture the Hamiltonian of the system bath interaction can be presented as,
\begin{eqnarray}
H_{SB}(t)&=&\sum_{\omega}A^{21\dag}(\omega)\otimes|1\rangle\langle 2|\otimes B(t)e^{i\omega t}\\\nonumber
& &+\sum_{\omega'}A^{12}(\omega')\otimes|1\rangle\langle 2|\otimes B(t)e^{-i\omega' t}+\mathrm{h.c.},
\end{eqnarray}
where $B(t)$ is
\begin{equation}
B(t)=e^{itH_B}B e^{-itH_B}=\sum_ng_na_ne^{-i\omega_nt}+g_n^*a_n^\dag e^{i\omega_nt}.
\end{equation}
The operators $A^{21\dag}(\omega)$ and $A^{12\dag}(\omega)$ are defined as,
\begin{equation}
A^{21\dag}(\omega)=\sum_{\lambda_i-\xi_j=\omega, \omega\geq0}|\lambda_i\rangle\langle \lambda_i|A|\xi_j\rangle\langle \xi_j|
\end{equation}
and
\begin{equation}
A^{12\dag}(\omega)=\sum_{\xi_i-\lambda_j=\omega, \omega\geq0}|\xi_i\rangle\langle \xi_i|A|\lambda_j\rangle\langle \lambda_j|.
\end{equation}

We assume that the walker is weakly interacting with the environment so that we can apply the Born-Markov approximation \cite{toqs}
\begin{equation}
\frac{d}{dt}\rho_s(t)=-\int_0^\infty ds \begin{mathrm}Tr\end{mathrm} _B\left[H_{SB}(t),\left[H_{SB}(t-s),\rho_s(t)\otimes\rho_B\right]\right],
\end{equation}
where $\rho_s(t)$ is the reduced density matrix of the system and $\rho_B$ is time independent state of the environment. In this paper we assume that the environment is in a thermal equilibrium state, i.e., the state of the bath is given by the canonical distribution $\rho_B=\exp{(-\beta H_B)}/\begin{mathrm}Tr\end{mathrm}[\exp{(-\beta H_B)}]$, where $\beta$ is the inverse temperature of the bath $\beta=(k_BT)^{-1}$.
Using the explicit form of the system-bath interaction Hamiltonian in the interaction picture and utilising the rotating wave approximation for the transition frequencies $\omega$ we obtain,
\begin{eqnarray}
\label{eq:DM}
\frac{d}{dt}\rho_s(t)&=&\sum_{\omega}\gamma(-\omega)\begin{mathcal}D\end{mathcal}\left(A^{21}(\omega)\otimes|2\rangle\langle 1|\right)\rho_s(t)\\\nonumber
& &+\gamma(\omega)\begin{mathcal}D\end{mathcal}\left(A^{21\dag}(\omega)\otimes|1\rangle\langle 2|\right)\rho_s(t)\\\nonumber
& &+\sum_{\omega'}\gamma(-\omega')\begin{mathcal}D\end{mathcal}\left(A^{12}(\omega')\otimes|1\rangle\langle 2|\right)\rho_s(t)\\\nonumber
& &+\gamma(\omega')\begin{mathcal}D\end{mathcal}\left(A^{12\dag}(\omega')\otimes|2\rangle\langle 1|\right)\rho_s(t)\nonumber,
\end{eqnarray}
where $\begin{mathcal}D\end{mathcal}(M)\rho$ denotes standard dissipative superoperator in a GKSL-form \cite{GKS,lin},
\begin{equation}
\begin{mathcal}D\end{mathcal}(M)\rho=M\rho M^{\dag}-\frac{1}{2}\left\{M^\dag M,\rho\right\}
\end{equation}
and $\gamma(\omega)$ is the real part of the Fourier transform of the bath correlation functions $\langle B^\dag(s)B(0)\rangle$,
\begin{equation}
\gamma(\pm\omega)=\frac{\gamma_0}{2}\left(\coth\left(\frac{\beta\omega}{2}\right)\mp1\right),
\end{equation}
where $\gamma_0$ is the coefficient of the spontaneous emission. Please note that in Eq. (\ref{eq:DM}) we drop the Lamb-shift type term. This term describes a shift of energy levels of the system due to the interaction with the environment and does not affect dissipative dynamics. Typically, the value of this Lamb-type shifts is few orders of magnitude smaller than the other characteristic parameters of the system Hamiltonian and traditionally this term is neglected.  

The quantum master equation (\ref{eq:DM}) has the form of the generalised master equation introduced by Breuer \cite{breuer}. It is easy to see that the following form of the density matrix 
\begin{equation}
\rho_s(t)=\rho_1(t)\otimes|1\rangle\langle 1|+\rho_2(t)\otimes|2\rangle\langle 2|
\end{equation}
is conserved. By the direct substitution to the quantum master equation (\ref{eq:DM}) we obtain,
\begin{eqnarray}
\label{eq:syst}
\frac{d}{dt}\rho_1(t)&=&\sum_\omega \gamma(\omega)A^{21\dag}(\omega)\rho_2(t)A^{21}(\omega)\\\nonumber
& &-\frac{\gamma(-\omega)}{2}\left\{A^{21\dag}(\omega)A^{21}(\omega),\rho_1(t)\right\}\\\nonumber
& &+\sum_{\omega'} \gamma(-\omega')A^{12}(\omega')\rho_2(t)A^{12\dag}(\omega')\\\nonumber
& &-\frac{\gamma(\omega')}{2}\left\{A^{12}(\omega')A^{12\dag}(\omega'),\rho_1(t)\right\}.\\\nonumber
\frac{d}{dt}\rho_2(t)&=&\sum_\omega \gamma(-\omega)A^{21}(\omega)\rho_1(t)A^{21\dag}(\omega)\\\nonumber
& &-\frac{\gamma(\omega)}{2}\left\{A^{21}(\omega)A^{21\dag}(\omega),\rho_2(t)\right\}\\\nonumber
& &+\sum_{\omega'} \gamma(\omega')A^{12\dag}(\omega')\rho_1(t)A^{12}(\omega')\\\nonumber
& &-\frac{\gamma(-\omega')}{2}\left\{A^{12\dag}(\omega')A^{12}(\omega'),\rho_2(t)\right\},\\\nonumber
\end{eqnarray}
This system of differential equations defines a continuous time version of the open quantum walk.

\section{Construction of a discrete time open quantum walk}

There are at least two ways of constructing the explicit form of the transition operators $B_i^j$ . The first one is to solve the system of matrix differential equations (\ref{eq:syst}) and discretize the solution. The second way is in deriving transition operators directly from the system of equations (\ref{eq:syst}) for a infinitesimal time step $\Delta$. In this paper we follow the second way. By replacing the derivative of operators $\rho_i(t)$ by the finite difference
\begin{equation}
\frac{d}{dt}\rho_i(t)\rightarrow\frac{\rho_i(t+\Delta)-\rho_i(t)}{\Delta}
\end{equation}
we can introduce the transition operators in the following way,
\begin{eqnarray}
B_2^1(\omega)&=&\sqrt{\Delta\gamma(\omega)}A^{21\dag}(\omega), \quad B_2^1(\omega')=\sqrt{\Delta\gamma(-\omega')}A^{12}(\omega'),\\\nonumber
B_1^2(\omega)&=&\sqrt{\Delta\gamma(-\omega)}A^{21}(\omega), \quad B_1^2(\omega')=\sqrt{\Delta\gamma(\omega')}A^{12\dag}(\omega'),\\\nonumber
B_1^1&=&\left[I_N-\frac{\Delta}{2}\sum_\omega\gamma(-\omega)A^{21\dag}(\omega)A^{21}(\omega)-\frac{\Delta}{2}\sum_{\omega'}\gamma(\omega')A^{12}(\omega')A^{12\dag}(\omega')\right],\\\nonumber
B_2^2&=&\left[I_N-\frac{\Delta}{2}\sum_\omega\gamma(\omega)A^{21}(\omega)A^{21\dag}(\omega)-\frac{\Delta}{2}\sum_{\omega'}\gamma(-\omega')A^{12\dag}(\omega')A^{12}(\omega')\right],
\end{eqnarray}
where $I_N$ is an identity operator on the node Hilbert space $\cal{H}_N$. It is straightforward to see that up to $\begin{cal}O\end{cal}(\Delta^2)$ these transition operators satisfy the normalisation conditions Eqs. (\ref{eq:norm2}),
\begin{eqnarray}
B_1^{1\dag}B_1^1+\sum_\omega B_1^{2\dag}(\omega)B_1^2(\omega)+\sum_{\omega'} B_1^{2\dag}(\omega')B_1^2(\omega')=I_N,\\\nonumber
B_2^{2\dag}B_2^2+\sum_\omega B_2^{1\dag}(\omega)B_2^1(\omega)+\sum_{\omega'} B_2^{1\dag}(\omega')B_2^1(\omega')=I_N,
\end{eqnarray}
and the discrete time open quantum walk is obtained using the following iteration procedure
\begin{eqnarray}
\rho_1^{[n]}=B_1^1\rho_1^{[n-1]}B_1^{1\dag}+\sum_\omega B_2^1(\omega)\rho_2^{[n-1]}B_2^{1\dag}(\omega)+\sum_{\omega'} B_2^1(\omega')\rho_2^{[n-1]}B_2^{1\dag}(\omega'),\\\nonumber
\rho_2^{[n]}=B_2^2\rho_2^{[n-1]}B_2^{2\dag}+\sum_\omega B_1^2(\omega)\rho_1^{[n-1]}B_1^{2\dag}(\omega)+\sum_{\omega'} B_1^2(\omega')\rho_1^{[n-1]}B_1^{2\dag}(\omega').
\end{eqnarray}

\section{Open quantum walk of the two-level system on the two-node graph}

Let us demonstrate the derivation of OQW on the simple example of the two-level system on the two-node graph. The Hamiltonian of the system $H_S$ reads,
\begin{equation}
H_S=\frac{\omega_0}{2}\sigma_z\otimes|1\rangle\langle 1|+\frac{\omega_0}{2}\sigma_z\otimes|2\rangle\langle 2|.
\end{equation}
The Hamiltonian of system-bath interaction can be written as
\begin{equation}
H_{SB}=\sum_n g_n \sigma_+\otimes|2\rangle\langle 1|\otimes a_n+\mathrm{h.c.}.
\end{equation}
The Hamiltonian of the bath is given by Eq. (\ref{eq:bath}). In this case there is only one possible transition with the transition operator $A^{21}=\sigma_-$ and the corresponding frequency of the transition is given by $\omega_0$.
The system of differential equations (\ref{eq:syst}) reduces to
\begin{eqnarray}
\frac{d}{dt}\rho_1(t)&=&\gamma(-\omega_0)\sigma_-\rho_2(t)\sigma_+-\frac{\gamma(\omega_0)}{2}\left\{\sigma_-\sigma_+,\rho_1(t)\right\},\\\nonumber
\frac{d}{dt}\rho_2(t)&=&\gamma(\omega_0)\sigma_+\rho_1(t)\sigma_--\frac{\gamma(-\omega_0)}{2}\left\{\sigma_+\sigma_-,\rho_2(t)\right\}.
\end{eqnarray}
For this system the transition operators $B_i^j$ of the open quantum walk take the form
\begin{eqnarray}
B_2^1&=&\sqrt{\Delta\gamma(-\omega_0)}\sigma_-,\quad B_1^2 =\sqrt{\Delta\gamma(\omega_0)}\sigma_+,\\\nonumber
B_1^1&=&\left(I_2-\frac{\Delta\gamma(\omega_0)}{2}\sigma_-\sigma_+\right),\quad B_2^2=\left(I_2-\frac{\Delta\gamma(-\omega_0)}{2}\sigma_+\sigma_-\right).
\end{eqnarray}
The dynamics of the probability to find the walker on one of the nodes is presented in Fig. 2. One can see that initially the probability to find the walker in the node oscillates. However, after 10 steps of the walk, the system reaches the equilibrium state. 

\begin{figure}
\begin{center}
\includegraphics[width= .7\linewidth]{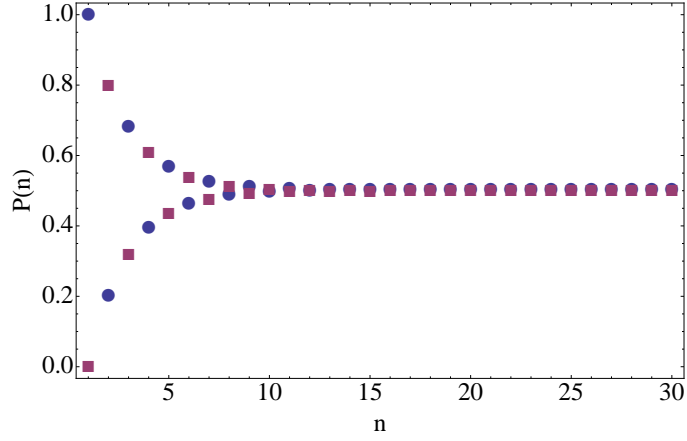}
\caption{The probability to find an open quantum walker in the node $1$ (circle) and in the node $2$ (square) as a function of the number of steps of the OQW. Initially, the walker was in the state $\rho^{[0]}=\sigma_-\sigma_+\otimes|1\rangle\langle 1|$. The other parameters of the model are chosen to be $\omega_0\beta=0.01$ and $\gamma_0\Delta=0.01$.}
\end{center}
\end{figure}

\section{Conclusion}

In this paper we introduced a microscopic model for the system environment interaction which leads to a discrete time open quantum walk on a two-node graph. We derive a quantum Markov master equation and show that after time discretization we obtain OQWs. We demonstrate the formalism for the example of the two-level atom on a two node graph. A numerical simulation shows the convergence of the walk to the stationary state.  

\section*{Acknowledgements}
This work is based upon research supported by the South African
Research Chair Initiative of the Department of Science and
Technology and National Research Foundation.

%This work is based on the research supported by the National Research Foundation

\end{document}